\documentclass[iop,apj,tighten]{emulateapj}
\usepackage{amsmath,amstext}
\usepackage{apjfonts}
\usepackage[breaklinks,colorlinks,citecolor=blue,linkcolor=magenta]{hyperref} 
\usepackage{float}
\usepackage{footmisc}
\usepackage{multirow}
\usepackage{lineno}

\def \nustar {{\em NuSTAR}}
\def \xmm {{\em XMM-Newton}}

\def \suzaku {{\em Suzaku}}

\def \swift {{\em Swift}}

\def \object {{MAXI J1820+070}}
\def \gx {{GX 339--4}}

\newcommand{\Msun}      {\mbox{$M_{\mathord\odot}$}}

\shortauthors{Xu et al.}

\begin{document}

\title{Evidence for Disk Truncation at Low Accretion States of the Black Hole Binary \\ MAXI J1820+070 Observed by \nustar\ and \xmm}

\author{Yanjun Xu\altaffilmark{1}}
\author{Fiona A. Harrison\altaffilmark{1}}
\author{John A. Tomsick\altaffilmark{2}}
\author{Jeremy Hare\altaffilmark{2,3,4}}
\author{Andrew C. Fabian\altaffilmark{5}}
\author{Dominic J. Walton\altaffilmark{5}}

\affil{$^{1}$ Cahill Center for Astronomy and Astrophysics, California Institute of Technology, Pasadena, CA 91125, USA}
\affil{$^{2}$ Space Sciences Laboratory, 7 Gauss Way, University of California, Berkeley, CA 94720-7450, USA}
\affil{$^{3}$ NASA Goddard Space Flight Center, Greenbelt,  MD 20771, USA}
\affil{$^{4}$ NASA Postdoctoral Program Fellow}
\affil{$^{5}$ Institute of Astronomy, University of Cambridge, Madingley Road, Cambridge CB3 0HA, UK}

\begin{abstract}
We present results from \nustar\ and \xmm\ observations of the new black hole X-ray binary \object\ at low accretion rates (below 1\% of the Eddington luminosity). We detect a narrow Fe K$\alpha$ emission line, in contrast to the broad and asymmetric Fe K$\alpha$ line profiles commonly present in black hole binaries at high accretion rates. The narrow line, with weak relativistic broadening, indicates that the Fe K$\alpha$ line is produced at a large disk radius. Fitting with disk reflection models assuming standard disk emissivity finds a large disk truncation radius (a few tens to a few hundreds of gravitational radii, depending on the disk inclination). In addition, we detect a quasi-periodic oscillation (QPO) varying in frequency between $11.6\pm0.2$~mHz and $2.8\pm0.1$~mHz. The very low QPO frequencies suggest a large size for the optically-thin Comptonization region according to the Lense-Thirring precession model, supporting that the accretion disk recedes from the ISCO and is replaced by advection-dominated accretion flow at low accretion rates. We also discuss the possibility of an alternative accretion geometry that the narrow Fe K$\alpha$ line is produced by a lamppost corona with a large height illuminating the disk.

\end{abstract}


\section{Introduction} \label{sec:intro}
Low-mass black hole X-ray binaries contain stellar-remnant black holes accreting from donor stars with a mass of $<1\Msun$ that transfer mass by Roche lobe overflow. Most known Galactic black hole X-ray binaries are low-mass X-ray binaries (LMXBs) and are discovered as transients. These systems exhibit recurrent outbursts on year to decade timescales, during which their flux increases by several orders of magnitude in the optical and X-ray bands \citep[e.g.,][]{corral16}. During a typical outburst, lasting for a few months, a black hole binary transitions between different X-ray spectral states, and displays distinct X-ray spectral and timing properties \citep[see][for a review]{rem06}.

\object\ (=ASASSN-18ey) is a new transient black hole X-ray binary discovered in 2018. The outburst was first reported in optical by the All-Sky Automated Survey for SuperNovae on UT 2018 March 06.58 \citep[ASAS-SN;][]{asassn_report}, and subsequently in the X-ray band by MAXI a week later \citep{maxi_report}. The source reached a peak X-ray luminosity (2--20~keV) of about 2 Crab, becoming one of the brightest X-ray nova discovered. Its outburst was well covered by multi-wavelength observations from the radio to the gamma-ray band \citep[e.g.,][]{bright18,shida19, paice19, hoang2019}. The distance of \object\ is estimated as $5.1\pm2.7$~kpc or $4.4\pm2.4$~kpc \citep{atri19}, and  3.46$^{+2.18}_{-1.03}$ kpc \citep{gandhi19} based on different distance priors from the {\em Gaia} DR2 parallax \citep{gaiadr2}. From radio parallax, the distance is measured to be $2.96\pm0.03$~kpc \citep{atri19}. \object\ is recently dynamically confirmed as a black hole binary (BHB) accreting from a K3-5 type donor star with a 0.68 day orbit, and the mass estimate of the central black hole is 7--8 \Msun\ \citep{maxi_1820_dyn}. Due to its high X-ray flux, \object\ is an ideal new target for the study of the inner accretion flow properties around black holes via X-ray spectral and timing analyses. X-ray timing analysis of {\em NICER} observations provides clues about the evolution in the coronal geometry \citep{kara19}. \nustar\ observations of \object\ during the bright phases of the hard state display relativistic disk reflection features, including a broad Fe K$\alpha$ line that is nearly invariant in the line profile over multiple observations at different fluxes. Detailed modeling of the reflection spectra reveals that the inner edge of the accretion disk remains stable at about 5 gravitational radii, which indicates that the central black hole is likely to have a low spin \citep{bharali19, buisson19}. In addition, a low-frequency QPO varying in frequency is detected in \object\ in both optical and X-ray bands \citep[e.g.,][]{yu18, buisson19}. 

Black hole accretion is generally studied in two regimes, cold accretion flows of optically-thick material at high mass accretion rates; and optically-thin hot accretion flows at low accretion rates. The latter is thought to contain an advection-dominated accretion flow (ADAF and its variants, see \citealt{yuan14} for a review). The ADAF model explains the hard X-ray emission from black holes accreting at low accretion rates, e.g., BHBs in quiescence and low/hard states, low luminosity AGNs (LLAGNs), and also Sgr A*. In BHBs, the accretion geometry is deduced to be in the form of an optically-thick accretion disk at large disk radii, which evaporates and is replaced by ADAF close to the black hole \citep[e.g.,][]{esin97}. The transitional radius, or the truncation radius of the optically-thick accretion disk, is predicted to increase with decreasing accretion rate \citep[e.g.,][]{meyer00,taam12}. 

Observationally, the inner radius of the accretion disk in BHBs can be measured by modeling the strength and temperature of the thermal disk emission component \citep{zhang97}, or modeling the profile of Fe K$\alpha$ emission line that originates from reflection of hard X-ray photons from the corona by the accretion disk \citep{fabian89}. The line profile becomes broad and asymmetric, with an extended red wing originating from the vicinity of the black hole due to combinational effects of Doppler shift, relativistic beaming and gravitational redshift \citep[see][for a review]{miller07}. Recent observations of several bright BHBs by \nustar\ reveal clearly broad Fe K$\alpha$ lines, indicating that the accretion disk extends all the way down to the ISCO in the bright hard state (BHS) \citep[e.g.,][]{miller15, xu18a, xu18b, buisson19}. In the low hard state (LHS), the detection of a narrow Fe K$\alpha$ line, the evidence for disk truncation, is usually hindered by the faintness of the targets, the weakness of the reflection features, and the limited instrumental spectral resolution. So far, narrow Fe K$\alpha$ emission has only been convincingly detected in the BHB candidate \gx\ from a long \suzaku\ observation taken in 2008 \citep{tom09}.

\section{Observation and Data Reduction}
We triggered \nustar\ \citep{nustar} and \xmm\ \citep{xmm} observations of \object\ during the second rebrightening period after the the end of its 2018 main outburst \citep[][]{xu2019_atel}. During this rebrightening period, the source stayed in the LHS at a low accretion rate (below 1\% of the Eddington limit), but reached an X-ray flux level high enough to enable the detection of the Fe K$\alpha$ emission line.  

\begin{figure}
\centering
\includegraphics[width=0.50\textwidth]{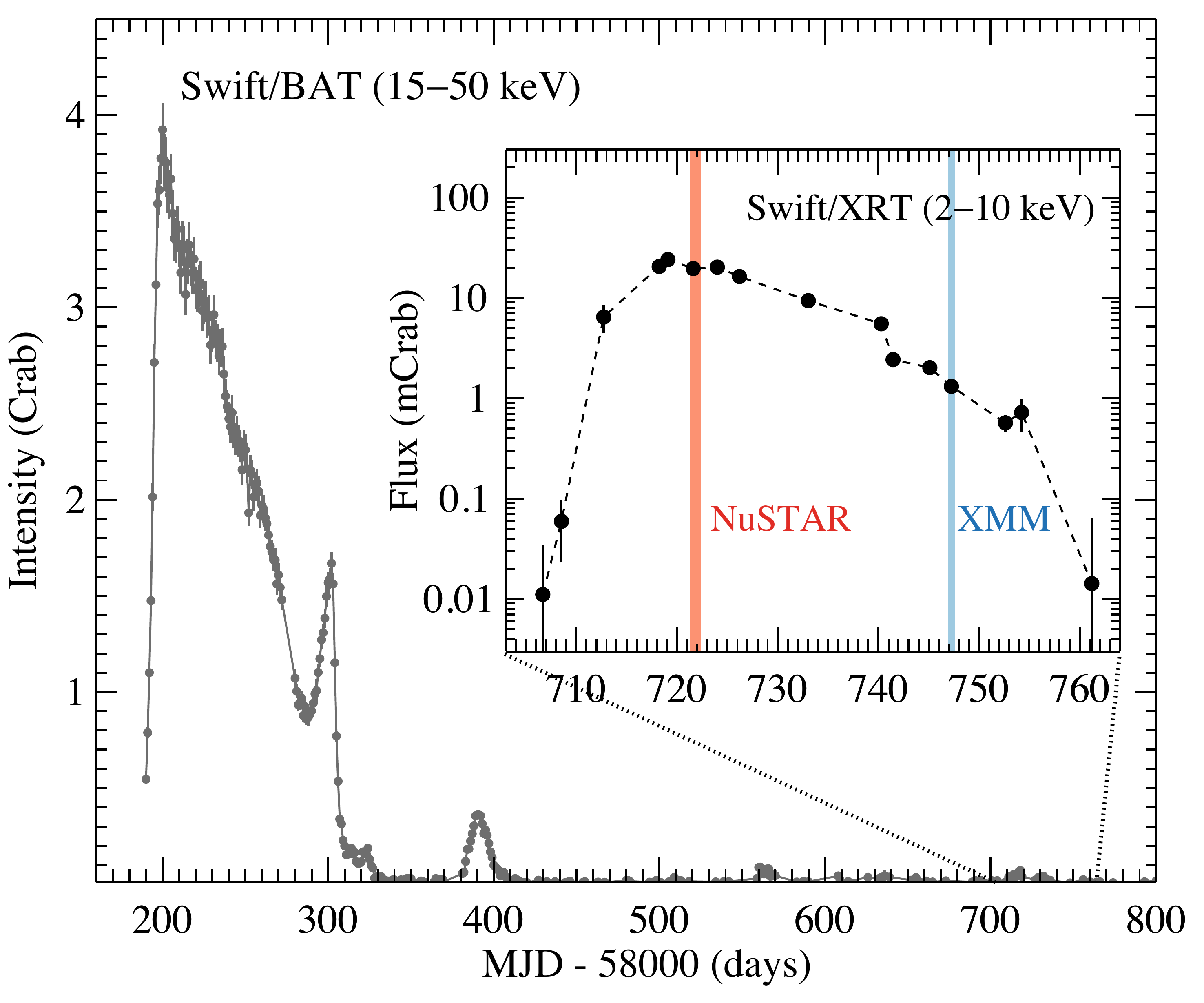}
\caption{Long-term \swift/BAT monitoring light curve of the outburst of \object. The BAT count rate is divided by the Crab count rate. The insert is the X-ray flux of \object\ measured by \swift/XRT close to the rebrighting period. The shaded areas mark the duration of the \nustar\ and \xmm\ observations analyzed in this work.
\label{fig:fig1}}
\end{figure}

We show the time of the \nustar\ and \xmm\ observations in the \swift/BAT \citep{swiftbat} and XRT \citep{swiftxrt} monitoring light curves of \object, see Figure~\ref{fig:fig1}. The \swift/BAT light curve was obtained from the \swift/BAT Hard X-ray Transient Monitor\footnote{https://swift.gsfc.nasa.gov/results/transients/} \citep{batmonitor}. We reduced the \swift/XRT data using {\tt xrtpipeline} v.0.13.5 with CALDB v20190910. We extracted source spectra from a circular region with a radius of 60\arcsec, and the background was extracted from an annulus with inner and outer radii of 160\arcsec and 300\arcsec. The X-ray flux in 2--10 keV measured by XRT was estimated by fitting the spectra in the energy range of 0.3--10~keV with an absorbed power-law model.

\subsection{NuSTAR}
\nustar\ observed \object\ on Aug 26, 2019 starting from UT 07:16:09 (ObsID: 90501337002). We reduced the data using NuSTARDAS pipeline v.1.8.0 and CALDB v20191008. The source spectra were extracted from a circular region with the radius of 180\arcsec\ from the two focal plane modules (FPMA and FPMB). Corresponding background spectra were extracted using polygonal regions from source-free areas. We also extracted spectra from mode 06 data to maximize the available exposure time following the method outlined in \cite{walton16}. The resulting exposure times are 48.6~ks and 48.4~ks for FPMA and FPMB, respectively. We coadded the FPMA and FPMB spectra using the {\tt addspec} tool in HEASOFT. The \nustar\ spectra were grouped to have a signal-to-noise ratio (S/N) of at least 10 per bin. We applied barycenter corrections to the event files, transferring the photon arrival times to the barycenter of the solar system using JPL Planetary Ephemeris DE-200, and extracted source light curve from the same region as the energy spectra.

\subsection{XMM-Newton}
The \xmm\ observation of \object\ (ObsID: 0851181301) started on September 20, 2019 from UT 22:45:46. Data reduction is performed using the \xmm\ Science Analysis System v17.0.0 following standard procedures. EPIC-MOS1, MOS2 and EPIC-pn operated in the timing mode, but EPIC-MOS1 experienced full scientific buffer during the whole observation. EPIC-pn \citep{struder01} is the prime instrument we use due to its large effective area in the Fe K band. The net exposure time of EPIC-pn is 56~ks after filtering out periods of high background flaring activity. The data is free from pile-up effects at a mean count rate of $\sim$20 cnts s$^{-1}$. We selected events with pattern $\leqslant$ 4 (singles and doubles) and quality flag = 0. The source spectrum and light curve were extracted from the columns of 27 $\leqslant$ RAWX $\leqslant$ 47, and the corresponding background was extracted from 58 $\leqslant$ RAWX $\leqslant$ 60. We used {\tt rmfgen} and {\tt arfgen} to generate the redistribution matrix files and ancillary response files. We grouped the EPIC-pn spectrum to have a minimum S/N of 10 for spectral modeling. The collected source light curve was barycenter corrected using the {\tt barycen} tool, and corrected for instrumental effects by {\tt epiclccorr}.

\section{Spectral Analysis}
In this work, we perform all spectral modeling in XSPEC v12.10.1f \citep{xspec}. We use the cross-sections from \cite{crosssec} and abundances from \cite{wil00} in the {\tt TBabs} neutral absorption model. All uncertainties are reported at the 90\% confidence level unless otherwise clarified. We fit \nustar\ spectra from 3.5--79 keV, and \xmm\ EPIC-pn spectrum from 0.6--10 keV.

\begin{figure*}
\centering
\includegraphics[width=\textwidth]{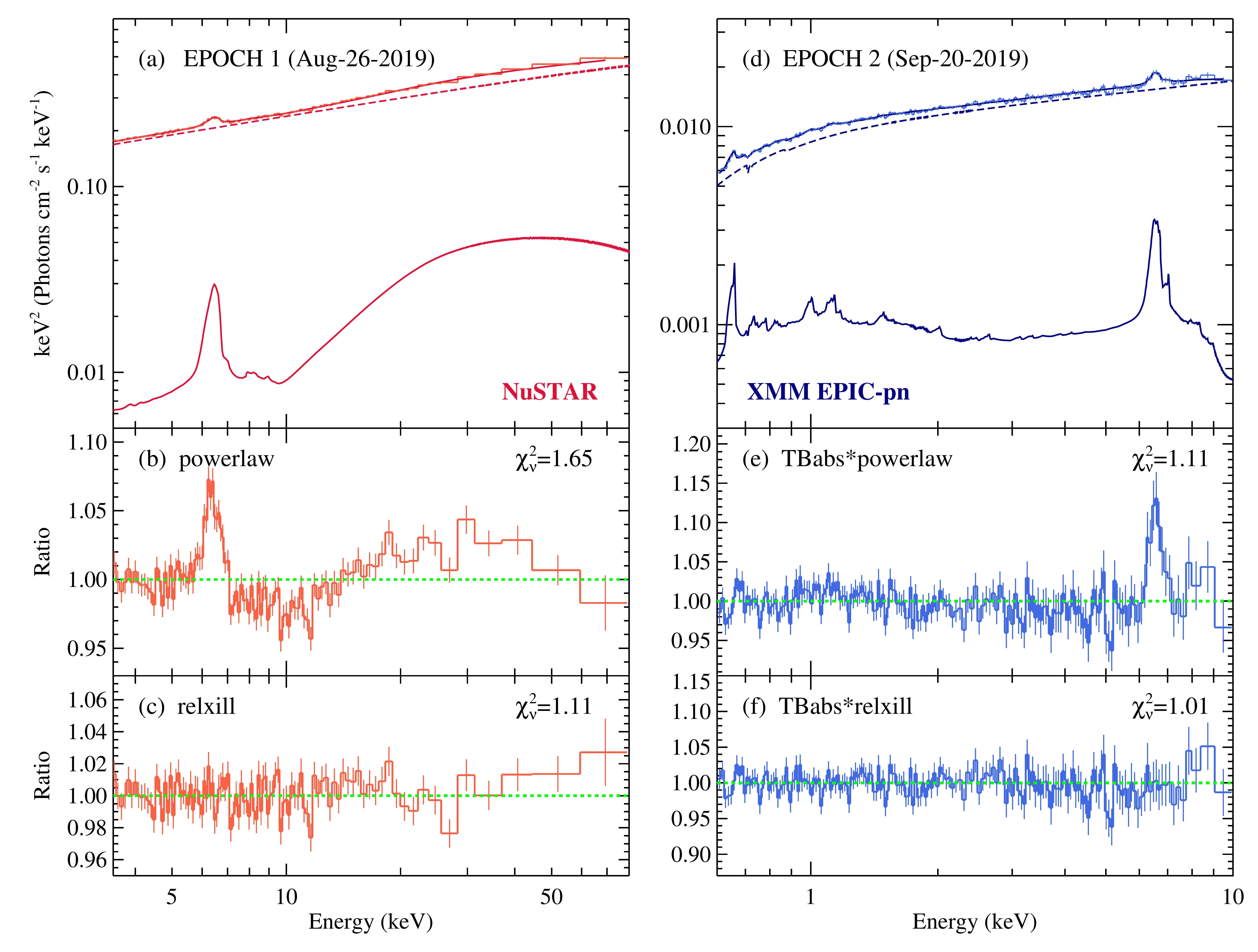}
\caption{Top: \nustar\ (EPOCH 1, left) and \xmm\ (EPOCH 2, right) spectra of the \object\ in the LHS, unfolded with the best-fit model. The coronal emission directly observed and the disk reflection component are plotted in dashed and solid lines, respectively. Middle and bottom: data/model spectral residuals. The spectra are rebinned for display clarity. The dips in the \nustar\ spectrum at about $11$~keV and $26$~keV and the excess in the \xmm\ EPIC-pn spectrum around 8~keV are calibration related. 
\label{fig:fig2}}
\end{figure*}

We first model the \nustar\ (=EPOCH 1) and \xmm\ (=EPOCH 2) spectra with an absorbed power-law model, {\tt TBabs*powerlaw}, in XSPEC. The \nustar\ and \xmm\ observations were taken in two epochs separated by 25 days, thus we fit the spectra individually with no linked parameters. We freeze the absorption column density, $N_{\rm H}$, at zero when modeling the \nustar\ spectra, as the small amount of absorption towards \object\ does not affect energies above 3~keV. 

\begin{deluxetable}{ccc}
\tablewidth{\columnwidth}
\tablecolumns{3}
\tabletypesize{\scriptsize}
\tablecaption{Energy Spectral Parameters of \object\ in the LHS \label{tab:tab1}}
\tablehead{
\colhead{Parameter} 
& \colhead{EPOCH 1}
& \colhead{EPOCH 2}
} 
\startdata
\multicolumn{3}{c}{Fe K$\alpha$ line modeled with {\tt Gaussian}: {\tt TBabs*(powerlaw+gauss)}} \\
\noalign{\smallskip} 
\hline
\noalign{\smallskip} 
$N_{\rm H}$ ($\rm \times10^{21}~cm^{-2}$)  &\nodata &$0.93\pm0.03$ \\
\noalign{\smallskip} 
E$_{\rm Fe}$~(keV) &$6.37\pm0.04$ &$6.54\pm0.06$ \\
\noalign{\smallskip} 
$\sigma_{\rm Fe}$~(keV) & $0.29^{+0.05}_{-0.04}$ &$0.24^{+0.09}_{-0.06}$ \\
\noalign{\smallskip} 
EW$_{\rm Fe}$~(eV) &$64\pm{7}$ &$98^{+20}_{-21}$ \\
\noalign{\smallskip}
$\Gamma$ &$1.667\pm0.006$ &$1.799\pm0.006$ \\
\noalign{\smallskip}
\hline
\noalign{\smallskip}
 $\chi^2/{\nu}$   &1173.4/939=1.25 &1160.3/1124=1.03   \\
\noalign{\smallskip}
\hline
\noalign{\smallskip} 
\multicolumn{3}{c}{Best-fit model:~{\tt TBabs*relxill}} \\
\noalign{\smallskip} 
\hline
\noalign{\smallskip} 
 $N_{\rm H}$ ($\rm \times10^{21}~cm^{-2}$)  &\nodata &{$1.03\pm0.04$}        \\
\noalign{\smallskip}                        
 $q$       &$3^f$    &$3^f$ \\
\noalign{\smallskip} 
 $\Gamma$      &$1.660^{+0.008}_{-0.007}$  &$1.765\pm{0.008}$                  \\
 \noalign{\smallskip}               
{$E_{\rm cut}$~(keV)}   &$>697$   &$>328$         \\
\noalign{\smallskip}
$i$ ($^\circ$)     &{$<20$}     &{$<39$}             \\
\noalign{\smallskip}
 $R_{\rm in}$ ($r_{\rm g}$)   &$27^{+10}_{-6}$     &{$>38$}              \\ 
\noalign{\smallskip}
 log~$({\xi})$ (log [$\rm erg~cm~s^{-1}$])  &$2.99^{+0.02}_{-0.17}$ &$3.08\pm0.06$                   \\
\noalign{\smallskip}                                                
 {$A_{\rm Fe}$ (solar)}     &$3.0^{+0.6}_{-0.5}$ &$6\pm2$                    \\
\noalign{\smallskip}                         $R_{\rm ref}$           &$0.060^{+0.005}_{-0.007}$  &$0.06\pm0.01$                    \\
\noalign{\smallskip}                                                
\hline                                                                                                        
\noalign{\smallskip}      $\chi^2/{\nu}$   &1033.8/936=1.10 &1135.4/1121=1.01   \\
\noalign{\smallskip}
\hline
\noalign{\smallskip}
$F_{\rm 2-10~keV}$~(erg~cm$^{-2}$~s$^{-1}$)  &$5.0\times10^{-10}$ &$3.9\times10^{-11}$ \\
$L_{\rm 0.1-100~keV}$~(erg~s$^{-1}$)  &$2.6\times10^{36}$ &$1.9\times10^{35}$ \\
\noalign{\smallskip} 
$L_{\rm Edd}$ (\%) &$0.25$ &$0.018$
\tablecomments{
Fixed parameters are marked with the superscript $f$. In this work, we adopt the distance estimate of 3~kpc and black hole mass estimate of 8 \Msun\ when calculating the source luminosity and Eddington ratio of \object.
}                               
\enddata  

\end{deluxetable}
\begin{figure}
\centering
\includegraphics[width=0.48\textwidth]{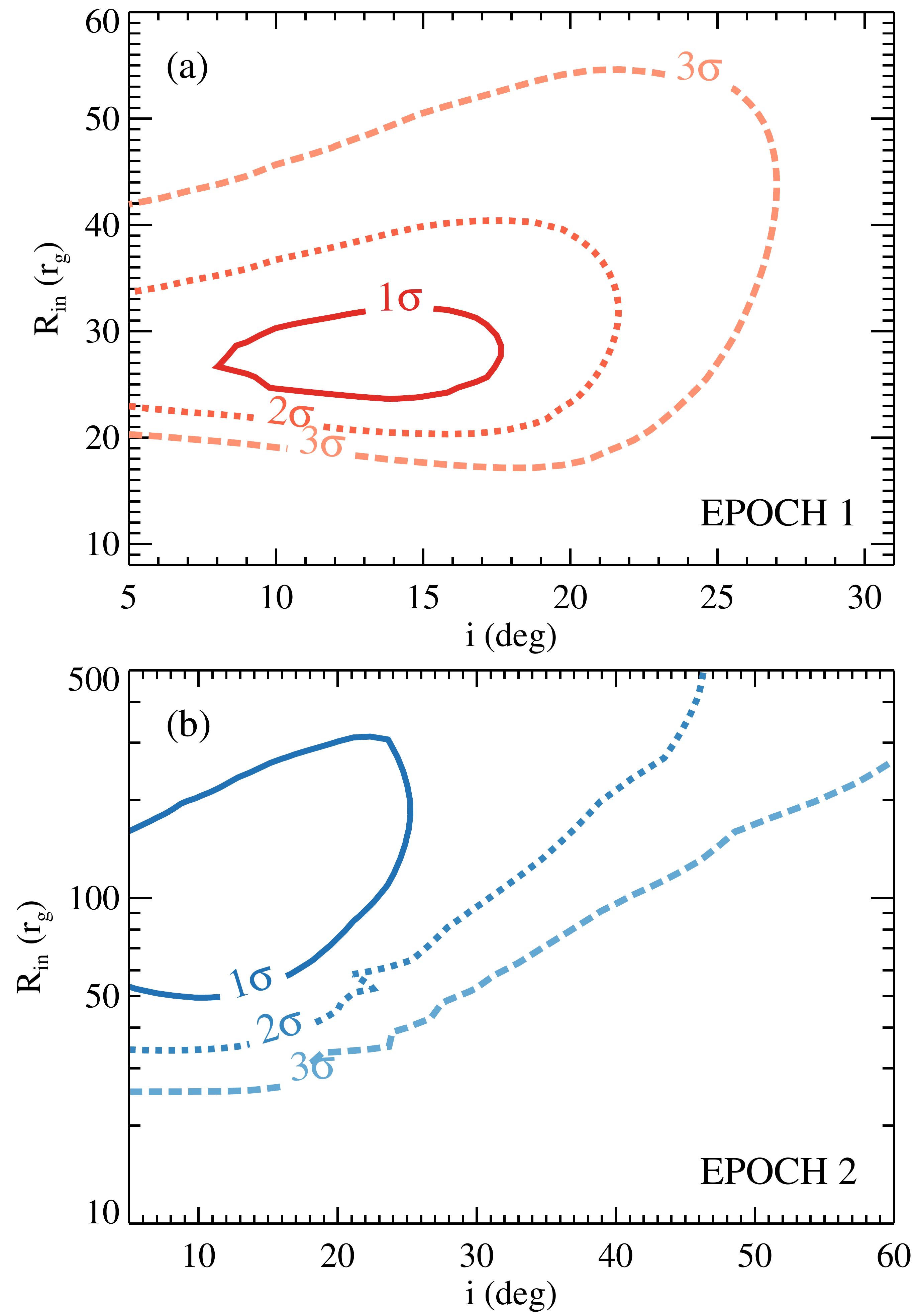}
\caption{Confidence contours of the inner disk radius, $R_{\rm in}$, and the disk inclination, $i$. 1$\sigma$, 2$\sigma$, and 3$\sigma$ corresponds to $\Delta\chi^2$ of 1, 4, and 9, respectively.
\label{fig:fig3}}
\end{figure}

The absorbed power-law model leaves systematic residuals, with a reduced chi-squared of $\chi^2/\nu~_{\rm EPOCH 1}=1557.1/941=1.65$ and $\chi^2/\nu~_{\rm EPOCH 2}=1248.3/1127=1.11$. 
Reflection features are evident in the spectral residuals (see Figure~\ref{fig:fig2}, panel (b) and (e)): an Fe K$\alpha$ emission line is clearly detected by both \nustar\ and \xmm\ centered around 6.4--6.5~keV, and a Compton reflection hump is evident in the \nustar\ spectra peaking around 30 keV. The prominent Compton reflection hump confirms that the Fe K$\alpha$ line originates from reflection by cold optically-thick gas \citep[e.g.,][]{lightman88}.  Accounting for the Fe K$\alpha$ line with a {\tt Gaussian} model improves the fit by $\Delta \chi^2=384$ for the \nustar\ spectrum, and $\Delta \chi^2=88$ for \xmm. The best-fit {\tt Gaussian} model indicates that the Fe K$\alpha$ line is narrow, with a line width of $\sigma_{\rm EPOCH1}=0.29^{+0.05}_{-0.04}$~keV measured by \nustar\ and $\sigma_{\rm EPOCH2}=0.24^{+0.09}_{-0.06}$~keV measured by \xmm, and equivalent width (EW) is constrained to be $64\pm7$~eV and $98^{+20}_{-21}$~eV (see Table~\ref{tab:tab1} for details). The line width and EW are significantly smaller than those measured with \xmm\ EPIC-pn during the BHS ($\sigma_{\rm BHS}\approx0.9$~keV and $EW_{\rm BHS}\approx270\pm30$~eV; \citealt{kaj19}). The narrow and symmetric line profile we observe here indicates weak relativistic effects, which is direct evidence for the Fe K$\alpha$ line being produced far from the central black hole.

In order to physically model the reflection features, and to get a constraint of the disk truncation radius, we fit the spectra with the {\tt relxill} relativistic disk reflection model ({\tt relxill} v1.3.3, \citealt{relxilla, relxillb}). We fix the disk emissivity indices, $q_{\rm in, out}$ at 3, the value expected for the outer part of a Shakura \& Sunyaev disk \citep[][]{laor91,dauser13}. This is a reasonable assumption, considering that the reflection features are produced at a large distance from the black hole. The emissivity indices cannot be constrained if left free. We fix the black hole spin, $a^*$, at the default value of 0.998. This parameter is irrelevant here as the inner disk is truncated outside of the ISCO. The outer disk radius, $R_{\rm out}$, is fixed at the maximum value of the model at 1000 $r_{\rm g}$ ($r_{\rm g}\equiv{\rm GM}/c^2$ is the gravitational radius). As the majority of the X-ray flux comes from the inner part of the accretion disk, the spectral modeling is not sensitive to $R_{\rm out}$. The {\tt relxill} model includes a coronal illuminating continuum in the shape of a power-law with an exponential cutoff at high energies, parameterized by the power-law index, $\Gamma$, and the high energy cutoff, $E_{\rm cut}$. Other free model parameters are the inner disk radius, $R_{\rm in}$, the disk inclination, $i$, the ionization parameter, $\xi$, the iron abundance, $A_{\rm Fe}$, and the reflection fraction, $R_{\rm ref}$.

The model, {\tt TBabs*relxill} in XSPEC notation, describes the data well, leaving no systematic structures in the residuals with $\chi^2/\nu_{\rm~EPOCH 1}=1033.8/936=1.10$ and $\chi^2/\nu_{\rm~EPOCH 2}=1135.4/1122=1.01$ (see Figure~\ref{fig:fig2}(c) and (f)). We measure a low absorption column density, $N_{\rm H}=(1.03\pm0.04)\times10^{21}$cm$^{-2}$ with \xmm, consistent with the value obtained early-on during the outburst \citep{shida19,kaj19}. The spectral continuum is well described by a power-law with a hard photon-index of $\Gamma\approx1.6-1.8$, with no prominent high energy cutoff required (see Table~\ref{tab:tab1} for best-fit parameters). The observed flux of \object\ in 2--10 keV decreased from  $5.0\times10^{-10}$~erg~cm$^{-2}$~s$^{-1}$ to $3.9\times10^{-11}$~erg~cm$^{-2}$~s$^{-1}$ between EPOCH 1 and EPOCH 2, corresponding to a change in Eddington ratio of $0.35$\% to $0.026$\%. Adding an extra thermal disk component modeled by {\tt diskbb} does not improve the fit, implying that thermal emission from the accretion disk is either too weak to be detected, or the disk is sufficiently cool that the peak of the disk blackbody distribution moves below the \xmm\ band.

The fitting results indicate that the disk is moderately ionized, with an ionization parameter of log$(\xi )\approx 3$. We measure the truncation radius of the optically-thick accretion disk to be $R_{\rm in,~EPOCH1}=27^{+10}_{-6}$~$r_{\rm g}$, and $R_{\rm in,~EPOCH2}>38$~$r_{\rm g}$. Based on the definition of the ionization parameter\footnote{$\xi$ is defined as $\xi= L_{\rm x}/(nr^2)$, where $L_{\rm x}$ is the X-ray luminosity, $r$ is the distance between the gas and the X-ray source, and $n$ is the gas density (hydrogen nucleus density).}, we estimate the density of the accretion disk, $n$, at the radius where the Fe K line is generated, decreases from $\sim10^{18}$~cm$^{-3}$ to $\sim10^{16}$~cm$^{-3}$ from EPOCH 1 to EPOCH 2. The best fit prefers a low inclination angle of the accretion disk, and only upper limits are obtained: $i_{\rm EPOCH1}<20^{\circ}$ and $i_{\rm EPOCH2}<39^{\circ}$. This is expected given the presence of a narrow Fe K$\alpha$ line, as line broadening caused by relativistic effects would become less apparent when the disk is viewed close to face-on \citep[][]{laor91}. $R_{\rm in}$ and $i$ are degenerate, as they are both related to the width of the Fe K$\alpha$ line. Without a more complicated line profile like the broad and asymmetric lines detected during the BHS, the two parameters cannot be uniquely constrained. To investigate the correlation between $R_{\rm in}$ and $i$, we plot the $\Delta\chi^2$ contour in Figure~\ref{fig:fig3}. As shown in panel (b), similar to case of \gx\ discussed in \cite{tom09}, the model tends to prefer a larger inner disk radius with increasing disk inclination. Therefore, we note that by letting both $R_{\rm in}$ and $i$ vary freely, we are quoting a conservatively small value for the disk truncation radius.

In addition, the best fit leads to a super-solar iron abundance, similar to that found in the BHS \citep{bharali19, buisson19}. It is currently uncertain whether the high iron abundance, frequently found when performing spectral modeling with ionized disk reflection models represents the true elemental abundance in the accretion disk, or is an overestimation resulting from physical processes overlooked in the calculation of the reflection models. There is evidence that this issue might be mitigated by using reflection models assuming high disk density \citep[e.g.,][]{tomsick18, jiang19}. We note that the iron abundance is known to be mostly related to the line strength rather than the line width, thus it is unlikely to have a significant effect on the estimate of the disk truncation radius here. 

The best-fit reflection fraction is $R_{\rm ref}\approx0.06$, significantly lower than that measured in the BHS of BHBs, which often requires a reflection fraction greater than unity. This reflection fraction in the BHS is believed to be enhanced by strong light-bending effects near the black hole \citep[e.g.,][]{mini04, reis13, xu18a, xu18b}. The reflection fraction parameter in the {\tt relxill} model is defined as the ratio of the coronal intensity illuminating the disk to that reaching the observer. The extremely low value we find here in the LHS of \object\ indicates that solid angle extended by the reflector is small, which is consistent with the scenario that the inner accretion disk is significantly truncated.

\section{Timing Analysis}
We produced the power spectral density (PSD) from the \nustar\ (EPOCH 1) and \xmm\ EPIC-pn (EPOCH 2) light curves, in the energy band of 3--79 keV and 0.6--10 keV. The \nustar\ light curves of FPMA and FPMB are added using the {\tt lcmath} tool in XRONOS. We produce the PSD from light curves with the time bins of 0.5~s, averaged in intervals of $2^{13}$ bins. The PSD is calculated in the rms normalization using {\tt powspec}, with white noise subtracted. The \nustar\ and \xmm\ PSD is geometrically rebinned by a factor of 1.03 and 1.05, respectively, to reach nearly equally spaced frequency bins in logarithmic scale. We fit the PSD in XSPEC with a multi-Lorentzian model using a unity response file: several zero-centered broad Lorentzians for the band-limited noise continuum, one narrow Lorentzian for the QPO and one for its possible sub-harmonic.

\begin{figure}
\centering
\includegraphics[width=0.48\textwidth]{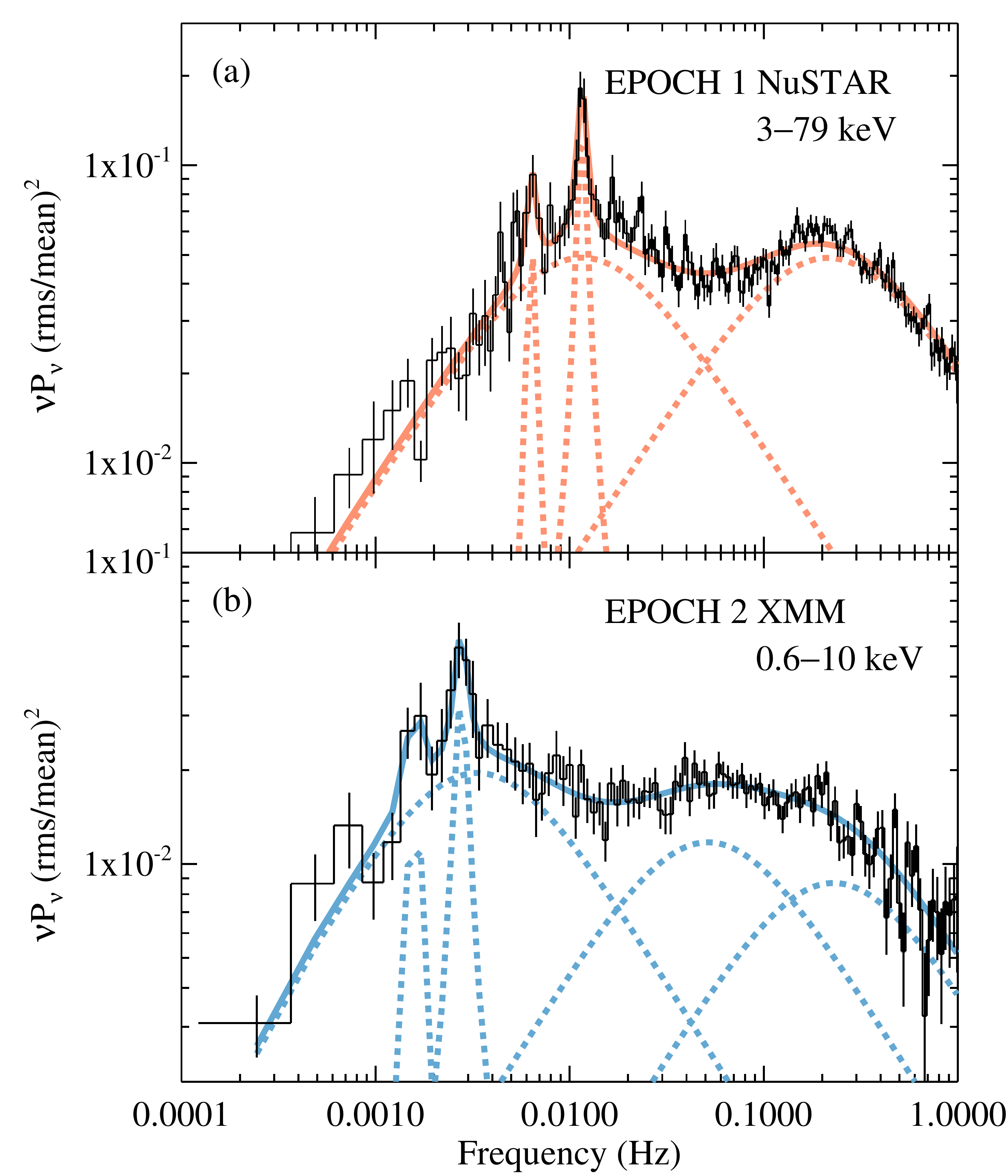}
\caption{\nustar\ and \xmm\ PSD of \object\ in the LHS. The best-fit models and their individual Lorentizian components are marked in solid and dashed lines.
\label{fig:fig4}}
\end{figure}

As shown in Figure~\ref{fig:fig4}, we find a QPO in the \nustar\ and \xmm\ PSD at the frequency of $\nu_{\rm EPOCH 1}=11.6\pm0.2$~mHz and $\nu_{\rm EPOCH 2}=2.8\pm0.1$~mHz, detected at $5.3\sigma$ and $3.2\sigma$ via the F-test. The QPO has rms variability of $13\pm1$\% and $11\pm3$\% in EPOCH 1 and EPOCH 2, respectively. 
The QPO is detected in the mHz range, lower than the typical frequency range of low-frequency QPOs found in BHBs (0.1--30 Hz). But the shape of the noise continuum and the fact that QPO is located close to the low-frequency break are consistent with type-C QPOs commonly found in BHBs \citep[][]{bell16}. The PSD is similar to that detected in \object\ during the BHS \citep{buisson19}, only with the QPO and the low-frequency break extending to even lower frequencies in the LHS.  

\begin{figure}
\centering
\includegraphics[width=0.48\textwidth]{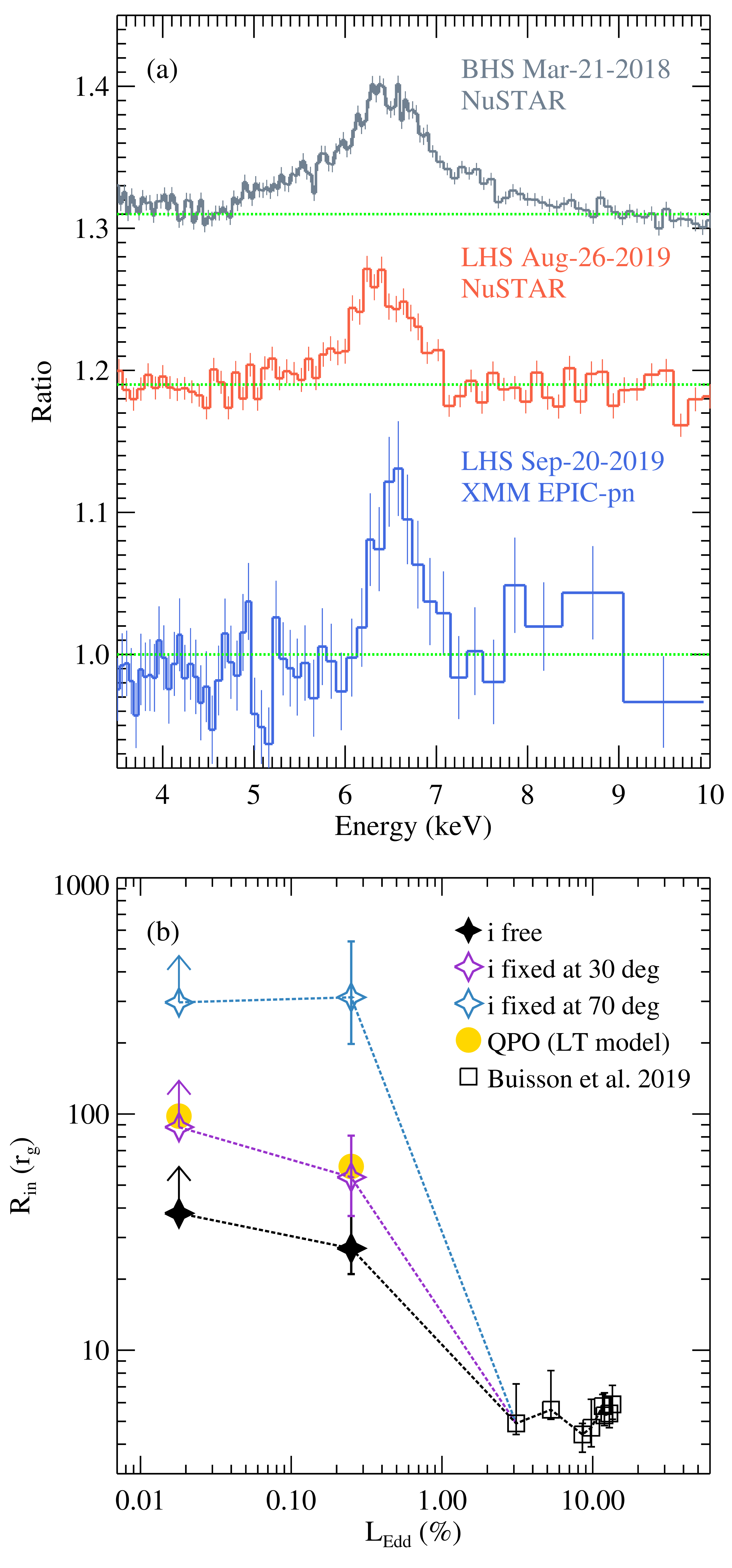}
\caption{(a) A comparison of the Fe K$\alpha$ line profile of \object\ observed at different epochs. The Fe K$\alpha$ becomes visually narrower during the LHS when compared that during the BHS \citep{buisson19}. We caution that the apparent difference in the narrow line profile observed by \nustar\ and \xmm\ during the LHS is at least partially due to the different energy resolution of the instruments, which is about 400 eV for \nustar\ and 100 eV for EPIC-pn on \xmm. (b) Evolution of the inner accretion disk radius with the Eddington ratio  measured in \object.
\label{fig:fig5}}
\end{figure}

\section{Discussion and Conclusion}
We detect a narrow Fe K$\alpha$ emission line with high S/N from \nustar\ and \xmm\ observations of the black hole X-ray binary \object\ during the second rebrightening period after its 2018 main outburst. The X-ray spectral and timing properties indicate that the source was in the LHS during the time of the observations. Spectral modeling reveals a very low absorption column density, combined with the moderate ionization state of the reflection material, confirming that the Fe K$\alpha$ line is produced from reflection by the accretion disk rather than that by torus-like Compton thick obscuring material commonly found in AGNs \citep{hickox18} and in the BHB V404 Cygni \citep{motta17}. The line is visibly narrow and lacks significant relativistic broadening, in contrast to the broad line profile observed during the BHS (see Figure \ref{fig:fig5}(a) for a comparison of the line profile\footnote{We choose the representative Fe K$\alpha$ line profile of \object\ in the BHS from the \nustar\ observation on March 21, 2018 (ObsID: 90401309006), detailed analysis of this dataset is published in \citealt{buisson19}.}), providing direct evidence for significant truncation of the inner accretion disk at low accretion rates in a BHB. 

There are disparities in the literature about the estimate of the disk inclination in \object. X-ray dips were observed during early phases of the outburst \citep{kaj19}, and a sharp increase in the H$\alpha$ emission line EW was reported and interpreted as a grazing eclipse of the accretion disk \citep{maxi_1820_dyn}, suggesting a high inclination of $i\approx60^{\circ}-80^{\circ}$ for the outer part of the accretion disk. We note that the inclination of this system is also unlikely to be very low (e.g, $< 10^{\circ}$) because radial velocity (RV) measurements are significant in amplitude \citep{maxi_1820_dyn}. The measured jet inclination angle of $62\pm3^{\circ}$ also indicates that the system is viewed at high inclination \citep{atri19}. In contrast, modeling the relativistic reflection spectra during the BHS yields a low inclination of $i\approx30^{\circ}$ for the inner part of the accretion disk \citep{buisson19, bharali19}. If both inclination estimates are robust, this implies a strong disk warp of $\sim30^{\circ}-50^{\circ}$. Our spectral fitting of the narrow Fe K$\alpha$ line in the LHS also prefers a very low inclination. However, as discussed above, it is expected for the spectral modeling to bias towards low inclinations in the presence of a narrow line profile. Without strong relativistic distortion effects, the inclination is poorly constrained as only upper limits are obtained. Therefore, we tried fitting the spectra with a fixed disk inclination angle instead, and still assuming a disk emissivity index of $q=3$. This results in a larger disk truncation radius and slightly degraded quality of the fits: when $i$ is frozen at 30$^{\circ}$, we get $R_{\rm in, 
~EPOCH1}=54^{+27}_{-17}~r_{\rm g}$ and $R_{\rm in, 
~EPOCH2}>88~r_{\rm g}$ with $\Delta\chi^2_{\rm EPOCH1}=11.7$ and $\Delta\chi^2_{\rm EPOCH2}=1.6$; when $i$ is fixed at $70^{\circ}$, the constraint on the inner disk radius becomes $R_{\rm in, 
~EPOCH1}=312^{+226}_{-114}~r_{\rm g}$ and $R_{\rm in, 
~EPOCH2}>297~r_{\rm g}$ with $\Delta\chi^2_{\rm EPOCH1}=9.9$ and $\Delta\chi^2_{\rm EPOCH2}=5.2$. These fits, although statistically slightly worse, leave no clear residuals with physical implications, thus may still be considered acceptable within calibration uncertainties.

There have been a number of observational campaigns aiming at investigating the evolution of the inner disk radius with the accretion rate in BHBs via the reflection method \citep[e.g.,][]{pet14, fuerst16}. However, the results are often poorly constrained or highly model dependent due to low statistics, especially at low flux states. In this work, we find evidence for a large inner disk radius in \object\ at low accretion rates (a few tens to a few hundreds of $r_{\rm g}$, depending on the disk inclination). Combined with earlier measurements during the BHS by \nustar\ \citep[][]{buisson19,bharali19}, it suggests that the inner edge of the accretion disk remains stable around the ISCO when the accretion rate is high, and starts to recede from the ISCO as the luminosity drops below $\sim1$\% of the Eddington luminosity (see Figure~\ref{fig:fig5}(b)). The critical accretion rate when significant disk truncation occurs is consistent with that measured in the well studied source \gx\ via the reflection method \citep[][]{tom09}, and that found based on the study of several BHBs by systematically modeling their thermal disk components \citep[][]{caba09}. In terms of the disk recessing with accretion rate, the results agree well with the theoretical prediction that the inner part of accretion disk becomes replaced by ADAF at low accretion rates (< 1\% of the Eddington limit; \citealt{esin97}, although the model predicts that the optically-thick accretion disk should be truncated in all hard states. 

In addition, we a detected QPO at $\nu_{\rm EPOCH 1}=11.6\pm0.2$~mHz and $\nu_{\rm EPOCH 2}=2.8\pm0.1$~mHz. The QPO and low-frequency break are found at lower frequencies than those in the BHS. Qualitatively, the longer time scales imply a physically larger size for the hot optically-thin Comptonization regions around black holes, where the QPO is believed to be generated. In the propagating mass accretion rate fluctuations model, the low-frequency break marks the viscous timescale at the outer edge of the Comptonization region \citep[e.g.,][]{ingram10, ingram13}. There have been various theoretical models put forward to explain the low-frequency QPOs in BHBs, but the exact mechanism is still highly uncertain. One of the currently promising models is the Lense-Thirring (LT) precession model. Adopting the simplified assumption that the QPO is caused by the effect of the Lense–Thirring precession of a test particle orbiting a spinning black hole at the disk truncation radius \citep{motta18}, we calculate the inner disk radius inferred from the QPO frequencies using the black hole mass of 8 \Msun\ and the spin of $a^*=0.3$. This leads to the characteristic truncation radius of $R_{\rm in,~EPOCH 1}\sim60$~$r_{\rm g}$ and $R_{\rm in,~EPOCH 2}\sim100$~$r_{\rm g}$. As a crude estimate, these are broadly similar to the spectral modeling results, in support of the inner accretion disk being significantly truncated in the LHS. During the BHS, however, we note that there is usually disagreement in the measurements from the two methods \citep[e.g.,][]{fuerst16, xu17, buisson19}, especially, \cite{buisson19} find that the QPO frequency and the disc inner radius are not connected. As discussed in \cite{ingram11}, it is possible that the discrepancy is related to physical complexities currently not well understood and thus not included in the QPO models, or complexities related to intrinsic properties of the corona.

Additional uncertainties in the measurement of the inner disk radius come from the poorly known nature of the corona, which affects the illumination pattern of disk \citep{fabian14}. During the above analysis, we assume the disk emissivity of $\epsilon \varpropto r^{-q}$ ($q=3$), which is expected for a standard accretion disk in the Newtonian regime. An alternative explanation for the narrow Fe K line that does not involve disk truncation is a low disk emissivity index ($q<2$), so that most of the contribution to the reflection features come from the outer disk. One possible accretion geometry that yields such a low emissivity index is a large lamppost height for the corona, which is believed to be associated with the base of a jet \citep[][]{dauser13}. The emissivity expected for a lamppost geometry in Newtonian gravity is $q\sim0$ for $r<h$, and $q\sim3$ for $r>h$ \citep{vaughan04}. The spectra can be equally well fitted by the reflection model assuming a lamppost geometry, {\tt relxilllp}, with a lamppost height of $h_{
\rm EPOCH 1}\sim40$~$r_{\rm g}$ and $h_{
\rm EPOCH 2}>70$~$r_{\rm g}$,  without the need to invoke disk truncation. But the model cannot self-consistently explain the low reflection fraction, unless significant beaming away from the accretion disk is involved. For a reflection fraction of $\sim0.1$, it requires bulk motion with a Lorentz factor of $\gamma\sim1.2$ when viewed at the inclination of 30$^{\circ}$, and $\gamma\sim1.6$ at the inclination of 60$^{\circ}$ \citep[][]{belo99}. It is uncertain whether strong beaming and the accretion geometry of a compact corona with a large lamppost height above the accretion disk are realistic descriptions of the system. There is evidence that significant beaming is absent in X-ray emission of BHBs in the LHS \citep[][]{nara05}. Although the case at low accretion rates is less clear, previous successful applications of the lamppost model to X-ray spectra of bright BHBs and AGNs measures a low lamppost height of $<10$~$r_{\rm g}$, or a steep disk emissivity profile \citep{fabian15}. Thus the physical implications of such a large lamppost height or low emissivity index required by the spectral fitting here are currently unclear and requires further investigation. The extended disk-corona model and the lamppost corona model are two competing coronal geometries that have been proposed \citep[][]{chau18}. The QPO frequency and low-frequency break in the PSD suggest that a physically large size for the Comptonization region, consistent with the extended corona and receding accretion disk scenario. The inner disk being truncated at a large radius also naturally
explains the non-detection of any thermal disk emission. Disk truncation provides a straightforward and physically reasonable explanation for the narrow Fe K$\alpha$ line we detected in \object. However, we note that an alternative accretion geometry of a high lamppost corona cannot be ruled out by our dataset and may also be plausible.

\acknowledgments{
We thank the anonymous referee for helpful comments that improved the paper. We thank Norbert Schartel for approving the \xmm\ DDT observation and the \xmm\ SOC for prompt scheduling of the observation. JH acknowledges support from an appointment to the NASA Postdoctoral Program at the Goddard Space Flight Center, administered by the USRA through a contract with NASA. DJW acknowledge support from an STFC Ernest Rutherford Fellowship. This work was supported under NASA contract No.~NNG08FD60C and made use of data from the \nustar\ mission, a project led by the California Institute of Technology, managed by the Jet Propulsion Laboratory, and funded by the National Aeronautics and Space Administration. We thank the \nustar\ Operations, Software, and Calibration teams for support with the execution and analysis of these observations. }

\bibliographystyle{yahapj}
\bibliography{maxi_1820_low_flux.bib}
\end{document}